# Magnetic field tuning of polaron losses in Fe doped $BaTiO_3$ single crystals

R. Anand Theerthan, Alla Artemenko and Mario Maglione

*ICMCB, Université Bordeaux, CNRS 87 avenue du Dr. A. Schweitzer, 33600 Pessac, France*

Abstract:

Artificial tuning of dielectric parameters can result from interface conductivity in polycrystalline materials. In ferroelectric single crystals, it was already shown that ferroelectric domain walls can be the source of such artificial coupling. We show here that low temperature dielectric losses can be tuned by a dc magnetic field. Since such losses were previously ascribed to polaron relaxation we suggest this results from the interaction of hopping polarons with the magnetic field. The fact that this losses alteration has no counterpart on the real part of the dielectric permittivity confirms that no interface is to be involved in this purely dynamical effect. The contribution of mobile charges hopping among Fe related centers was confirmed by ESR spectroscopy showing maximum intensity at *ca* T~ 40 K.

Introduction:

The quest for the novel electronic materials has revived the search for mutliferroic materials during the last decade in which more than one ferroic order coexist [1]. Although extensive research is been conducted in this field very few materials have shown to be promising, ex $BiFeO_3$, $TbMnO_3$ [2, 3]. Difficulty in finding mutliferroic materials stems from the fact that ferroelectric materials require cations to have empty 'd' orbitals whereas ferromagnetism occurs only in materials with partially filled 'd' orbitals making these two properties mutually exclusive [4]. Such difficulties lead researchers to look for artificial means of realizing magnetoelectric coupling, ex composite materials combining ferroelectrics and ferromagnets like $CoFe_2O_4$ and $BaTiO_3$ [5]. Recently Catalan has reported that apart from having magnetoelectric coupling, magnetoresistive artifacts can also invoke large magnetodielectric coupling, providing an additional means to stimulate Artificial Magneto Capacitance (AMC) [6]. According to his model, the interface between electrode and dielectric or grain boundaries in ceramics may have a resistance different from the core of the dielectric due to the difference in charge carrier density that give rise to Maxwell - Wagner effect and eventually to magnetodielectric effect caused by change of resistance at the interface under magnetic field. The essential ingredient in obtaining AMC is interfaces and free charges.

Recently Maglione has reported magnetocapacitance in one cent diodes as well as in $CaCu_3Ti_4O_{12}$ so called giant permittivity material [7]. Magnetocapacitance in diode stems from interaction of external magnetic field with free charges accumulated at the p-n interface. In case of $CaCu_3Ti_4O_{12}$ free charges arise from incomplete compensation of Cu related defects [8] and the interfaces are grain boundaries which under magnetic field generates magneto capacitance. In $BaTiO_3$, it was also shown that ferroelectric domain walls can be the necessary interfaces where to locate free charges (7). Here, we confirm similar artificial coupling in Fe doped barium titanate ($BaTiO_3$ - BTO) single crystals with different doping concentration where the interfaces are domain walls and the charged oxygen vacancies due to the substitution of $Fe^{3+}$ for $Ti^{4+}$ are the source of free charges. AMC is more pronounced at the temperature where the domain walls undergo relaxation (7). Unlike the grain boundaries which are rigid, domain walls can easily move under the influence of temperature and electric field which make the observed AMC unreliable. In contrast, we show here a reproducible tuning of dielectric losses in the same Fe doped BTO at low temperatures where the polarons due to trapping of electrons relaxes. While the losses in the relaxation range can be tuned by about 20%, no sizeable effect could be observed on the real part of the dielectric permittivity. We thus conclude that no interfaces are needed and that it is the direct interaction of the applied magnetic field with the hopping polarons which induces the overall losses change. These results show that AMC can be observed also in single crystals while it was previously reported in ceramics, polycrystalline films and bulk composites.

## Experimental Methods:

Single crystals of $BaTiO_3$ with two different concentrations of iron (0.075 at. % and 0.3 at. %) were investigated. Typical sample dimensions were 4 x 4 x 2 $mm^3$. Gold was sputtered on the major faces to make electrodes and silver wires glued to the center of the electrode with silver paste were used for electrical contact. The samples were then put in a Quantum Design Physical Properties Measurement System (PPMS) connected to four coaxial cables linked to an HP4194 impedance analyzer through BNC connectors. Capacitance and dielectric losses were measured as a function of temperature from room temperature down to 10 K in the frequency range of 100 Hz-10 MHz. For measurements under magnetic field, the field was raised at a rate of 200 $Oe.s^{-1}$ from 0 to 90 kOe; this field was then fixed throughout the temperature cycle. ESR measurements were performed using an X-band Bruker spectrometer operating at 9.4 GHz. An Oxford Instruments ESR 9 He cryostat operating in the temperature range 4 – 300 K was used for temperature dependence studies of ESR spectra intensities.

## Artificial dielectric modulation under a magnetic field:

Reliable measurement of dielectric parameters under a large magnetic field needs caution so as to avoid spurious results since even a small amount of magnetic impurities will give Artificial Magneto Capacitance (AMC) which may not have its origin in the sample. As a necessary precaution care should be taken to ensure that external factors (sample holders, connections etc) do not contribute to the observed relaxation especially at low temperatures. Calibration is thus required in the temperature range of interest. Hence a piece of Teflon was chosen first for open circuit calibration due to its very low dielectric losses. Fig 1(a) gives dielectric losses of Teflon as a function of temperature for a spot frequency of 590 kHz. Observed losses are indeed very low (less than 1%) and no relaxation or any other dielectric anomaly was noticed. Dielectric losses decrease continuously with decrease in temperature. As Teflon is a low permittivity material, we also performed calibration using single crystal of permittivity and losses similar to $BaTiO_3$: Fe, namely pure $KTaO_3$ and $SrTiO_3$. On figure 1(b), in agreement with former reports, we find a continuous increase of the capacitance and a slight increase of losses at about 50 K [9]. On this figure, the errors bars denote the long term isothermal evolution of the capacitance and losses when the magnetic field was swept from 0 to 90 kOe. We found the same absence of magnetic field effect on the low temperature dielectric properties of $SrTiO_3$ single crystals (not shown). This fixes the detection threshold of dielectric parameters under magnetic field to 3%.

Magnetic field effect on the dielectric relaxation of Fe doped BTO single crystals will be described in the following. Fig 2 (a) & (b) shows dielectric losses as a function of temperature for BTO doped with 0.075 at% and 0.3 at% Fe at a spot frequency of 10 KHz with 0 and 90 KOe (magnetic field) respectively. The sharp anomalies at 281 K and 198 K are phase transition from tetragonal to orthorhombic and orthorhombic to rhombohedral phases respectively. The transition temperatures are not the same as pure BTO due to Fe doping. The broad maximum noticed close to 155 K is due to domain wall relaxation where as the maximum at *ca* 30 K can be ascribed to polaronic relaxation arising from the presence of $Fe^{2+}$ and $Fe^{3+}$ ions [10, 11]. Low temperature polaron relaxation will be discussed in detail later; here we will focus on magnetic field effect on the high temperature relaxation. It is obvious from the fig 2 (a) that for BTO with 0.075 at% Fe, application of magnetic field has enhanced the amplitude at the relaxation maximum whereas for BTO with 0.3 at% Fe, magnetic field has depressed the amplitude of the relaxation (fig 2(b)). This shows that there is no systematic link between the effect of magnetic field and the amount of Fe. In addition, for a given Fe content, the AMC amplitude is changed on thermal cycling of the crystals. This confirms that domain walls which are highly sensitive to the sample thermal history are included in this AMC.

Now we look into the low temperature relaxation ascribed to polarons. Before going into detail we would like to stress again that it is important to confirm that the observed relaxation indeed stems from the sample. The experiment with Teflon which we have described earlier insures that the sample holder or other external factors do not contribute to the relaxation at low temperatures and it stems only from the sample. Fig 3(a) & (b) gives dielectric losses as a function of temperature for BTO doped with 0.075 at% and 0.3 at% at a spot frequency of 115 kHz with 0 and 90 KOe (magnetic field). The maximum of relaxation is around 30 K for both Fe doped crystals. Under the application of magnetic field of 90 kOe, amplitude of the relaxation is enhanced for BTO with 0.075% Fe and depressed for BTO with 0.3 at% Fe. In both cases, the temperature of the dielectric losses maximum was not shifted under magnetic field. This shows that the activation energy of 40 meV is not altered by the magnetic field. Only the magnitude of dielectric losses is affected. To appreciate the amplitude variation by magnetic field on this relaxation, dielectric losses was plotted as a function of frequency for BTO with 0.3 at% Fe at three different spot temperatures (20, 30 and 70 K) as shown in fig 4(a)-(c). It appears from fig 4(a) & (b) that the effect of magnetic field is more pronounced at 20 and 30 K respectively; it reaches 21% at a frequency of 10 kHz for 20 K where the maximum of polaronic relaxation occurs. At higher temperatures (70 K) sufficiently away from the relaxation maximum, the magnetic tuning of dielectric losses disappears (fig 4 (c)).

Microscopic investigation of the magnetic field effect on relaxation

In agreement with the Catalan model [6], it was previously shown that ferroelectric domain walls can act as interfaces leading to AMC [7]. Localization of free charges at such interfaces was held responsible for the tuning of macroscopic impedance versus the magnetic field. Having found AMC in Fe doped BTO, in this work we looked for dependency of AMC on doping concentration *i.e.* concentration of free charges. It is clear from fig 2 that artificial magneto capacitance observed shows no consistent trend with the doping concentration. Surprisingly the effect tends to vary between two different experiments of the same sample indicating non-reproducibility. This can be interpreted by means of kinetics of domain walls with temperature. The number of domain walls changes (i.e. interfaces) when the sample is subjected to several temperature cycles as it goes through phase transition many times. Since artificial magneto capacitance depends on both interfaces and charged defects, changing one of them would consequently change the magneto capacitance. We have no way of controlling the number of domain walls since it depends on the temperature, which makes the magnitude of this magneto capacitance effect highly irreproducible.

The low temperature dielectric relaxation is a more reliable way of achieving artificial magneto capacitance because this relaxation does not stem from domain wall dynamics. The same kind of dielectric losses anomaly arising from polarons was reported in perovskites around 40 K due to different valence states of dopants [9, 11]. In $BaTiO_3$ doped with Fe we found a relaxation at low temperatures in the temperature range of 20 to 50 K which is ascribed to polarons that appears due to existence of $Fe^{2+}$ and $Fe^{3+}$ in the crystal [12]. Our polaron model was further supported by the activation energy of 40 meV obtained for the relaxation which very well agrees with the previous results [11]. In fig 3 for BTO doped with 0.075 at% the magnetic field enhances the relaxation amplitude while it depresses for BTO doped with 0.3 at% Fe, showing contradiction. To understand the effect of magnetic field on the amplitude of this relaxation it is necessary to look deeply into the microscopic process happening in the crystal at low temperatures. The relaxation under discussion is due to hopping of free charges and it can happen in number of ways a.) by purely lattice related defects where there is a charge transfer between $Ti^{4+}$ and $Ti^{3+}$ centers or oxygen vacancy related centers b.) by extrinsic charged defects, in our case it is the hopping of free charges between different oxidation states of iron owing to doping of Fe . Such hopping process leads to dipoles reorientation which could then undergo relaxation. Importantly hopping of free charges induces a change in lattice elastic energy and the application of magnetic field does not affect this energy; this is confirmed by our finding of same activation energy for low temperature relaxation with 0 and 90KOe magnetic field. In contrast, magnetic field could affect the way these electrons are hopping provided the temperature and frequencies are favorable. This could explain why the magnetic field influences the amplitude of relaxation. It is immediately seen in Fig 4(a) & (b) where the magnetic field effect on the maximum of the dielectric losses is strongly marked at 20 K and 30 K along with high losses which points out the transfer of electron between different oxidation states of iron ions. On the other hand, at 70 K both no magnetic field effect and low dielectric losses were observed showing that the frequency and temperature are not favorable for hopping or polaron relaxation (fig 4(c)).

To find microscopic evidences of free charge hopping between Fe ions, we performed ESR studies in the nominally pure and the Fe-substituted $BaTiO_3$ single crystals. We point out that the crystals that were used for ESR studies are the same as the ones used for the dielectric experiments reported just above. Because of the very low detection limit of ESR and because of the unavoidable presence of iron in the nominally pure $BaTiO_3$, the ESR spectra recorded at 4 K look the same in all crystals with the most intensive resonance at g=2.0005 due to the central ±1/2 transition of $Fe^{3+}$ (fig 5); the lines' intensity is proportional to the iron concentration. The full analysis of these spectra including the rotation plots is out of the scope of the present paper. We only underline the $Fe^{3+}$ related lines and their behavior as a function of temperature. In agreement with previous reports, the line located at g=2.0005 (3350 Oe) is ascribed to $Fe^{3+}$ ($3d^5$ ion, with electron spin S=5/2) in the six-fold environment of oxygen. In the same way, the line at g=4.28 (1580 Oe), smaller in intensity, stems from $Fe^{3+}$ centers linked to 2 oxygen vacancies with random relative position. Because of their symmetry, these centers do not entail any change in the spectra on rotating the crystals versus the magnetic field. For both these lines, the intensity displays very distinct anomaly at *ca* 40 K. The VO-$Fe^{3+}$-VO line intensity decreases from 4 K and fully vanishes above 40 K ( fig 5, Inset (b)) and in contrast the six-fold oxygen coordination of $Fe^{3+}$ line grows from 20 K up to a large maximum at *ca* 40 K (fig 5, inset (a)). Both these features can be ascribed to electron exchanges between $Fe^{2+}$ and $Fe^{3+}$ centers. Keeping in mind that $Fe^{2+}$ ion ($3d^6$ ion, S=0) is not ESR active, then the decrease of the VO-$Fe^{3+}$-VO intensity can result from the electron delocalization from this charged $Fe^{3+}$ center. The disappearance of this center follows an Arrhenius law with activation energy of 0.13 eV which is in agreement with a shallow trap located close to the bottom of the conduction band. This was confirmed by magnetic measurement as a function of temperature in which the magnetization goes to negative above 30 K. On the other hand, we ascribe the 40 K maximum of the octahedral $Fe^{3+}$ center to the optimal electron exchanges among $Fe^{2+}$/$Fe^{3+}$ sites which reflects in the thermally activated losses. In such a case, the ESR maximum is signing the high electron hopping rate in this temperature range. More precisely, it is in this temperature range that the electron trapping on Fe centers is counter-balanced by the thermal activation of long range electronic motion. This also led to an optimal interaction of these hopping electrons with the applied magnetic field. Our temperature dependent ESR results are thus a strong support for the magnetic field induced losses without magneto-capacitance modulation.

Conclusion:

Artificial coupling of a magnetic field with dielectric parameters of ferroelectric single crystals can happen in several ways. Close to the ferroelectric domain wall relaxation range, in agreement with the Catalan model, a large artificial magneto capacitance effect has been confirmed in Fe doped $BaTiO_3$ crystals. However, due to the lack of control of the domain wall density, such effect is unreliable. In the low temperature range T<50 K, the magnetic field has no effect on the capacitance but it can tune the dielectric losses. This tuning can be more than 15% under 90 kOe in Fe-doped $BaTiO_3$ while it stays below the detection threshold in pure $KTaO_3$ and pure $SrTiO_3$. We suggest that such efficient losses tuning results from the interaction between the magnetic field and hopping polarons that increases the dielectric losses. When the density of such polarons is small like in pure crystals, no macroscopic effect can be measured. Unlike domain walls relaxation, this polaronic contribution does not need interfaces to happen which in turns explain why no magneto capacitance could be observed.

Acknowledgement

For ESR investigations financial support from a Marie Curie International Incoming Fellowship within the 7th European Community Framework Programme (GA 255662) is gratefully acknowledged.

**Figure Caption:**

**Fig 1:** (a) Temperature dependent dielectric losses for a Teflon sample at a spot frequency of 590 KHz. Losses are less than 1% and no dielectric anomaly was observed in the whole temperature range of measurement. (b) Capacitance and dielectric losses as a function of temperature for pure $KTaO_3$. Error bars denote the long term isothermal evolution of capacitance and losses when the magnetic field was raised from 0 to 90 KOe

**Fig 2:** Dielectric losses as a function of temperature under 0 and 90 KOe magnetic fields for a spot frequency of 10 KHz: (a) $BaTiO_3$ doped with 0.075 at% Fe, (b) $BaTiO_3$ doped with 0.3 at% Fe. Sharp anomalies at 281 K and 198 K represent phase transition from tetragonal to orthorhombic and orthorhombic to rhombohedral respectively.

**Fig 3:** Dielectric losses as a function of temperature under 0 and 90 KOe magnetic fields at a spot frequency of 115 KHz: (a) $BaTiO_3$ doped with 0.075 at% Fe, (b) $BaTiO_3$ doped with 0.3 at% Fe.

**Fig 4:** Dielectric losses as a function of frequency with 0 and 90 KOe magnetic field for BTO doped with 0.3 at% Fe at: (a) 20 K, (b) 30 K and (c) 70 K. Difference in amplitude was found with and without magnetic field in addition to high losses close to the relaxation maximum at 20 K and 30 K. Away from relaxation maximum, at 70 K no difference in amplitude was found under magnetic field, along with low losses was found. (Lines are drawn to guide the eyes)

**Fig 5:** ESR spectra of pure $BaTiO_3$ single crystals showing two lines, one at 3350 Oe (g=2.0005) and the other smaller at 1580 Oe (g=4.28). Both of them relate to $Fe^{3+}$ active centers with different environment (details in text). Inset (a) Temperature dependence of the integrated intensity for the line at 3350 Oe. Inset (b) Temperature dependence of intensity for the line at 1580 Oe

Figure 1

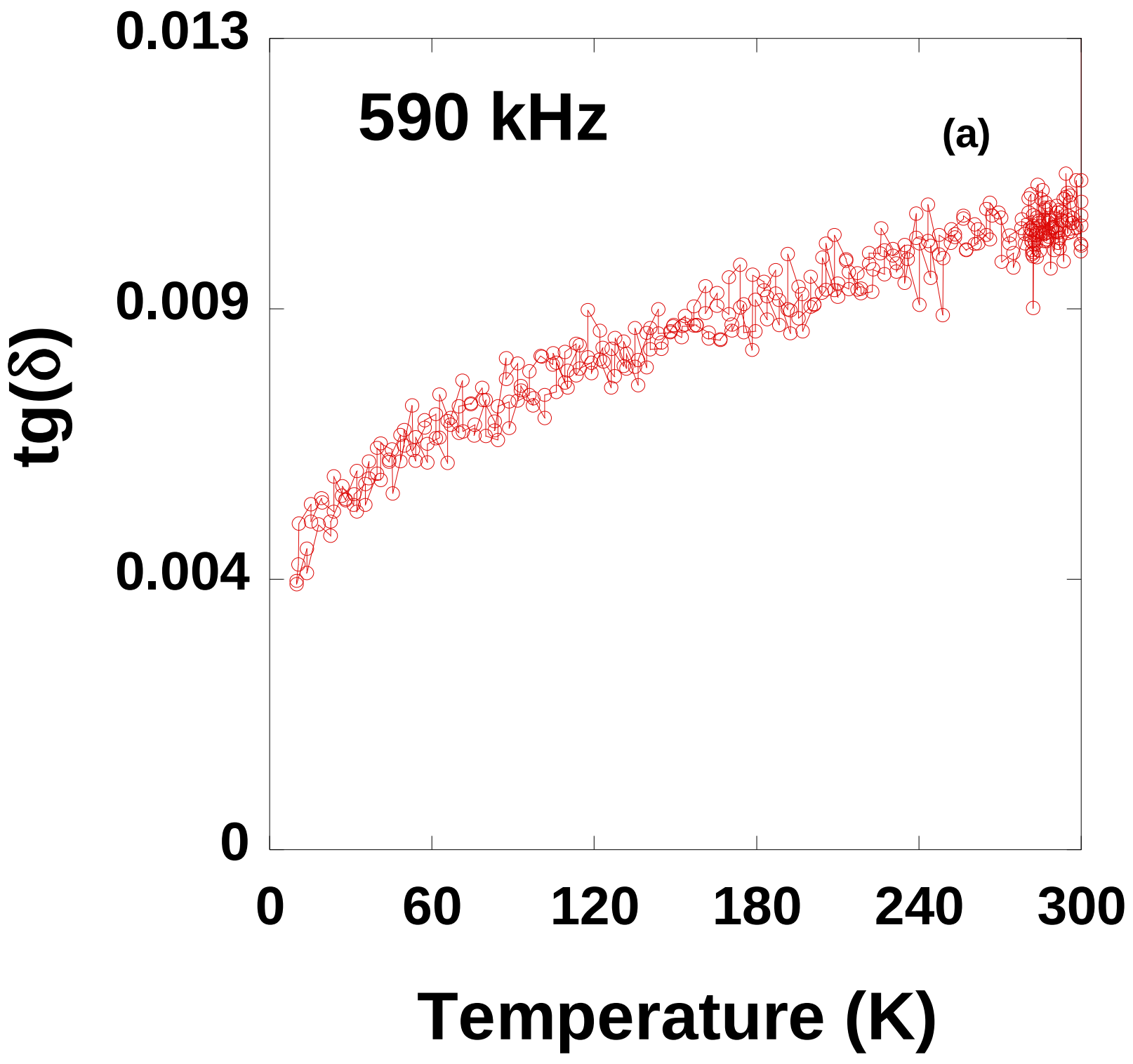

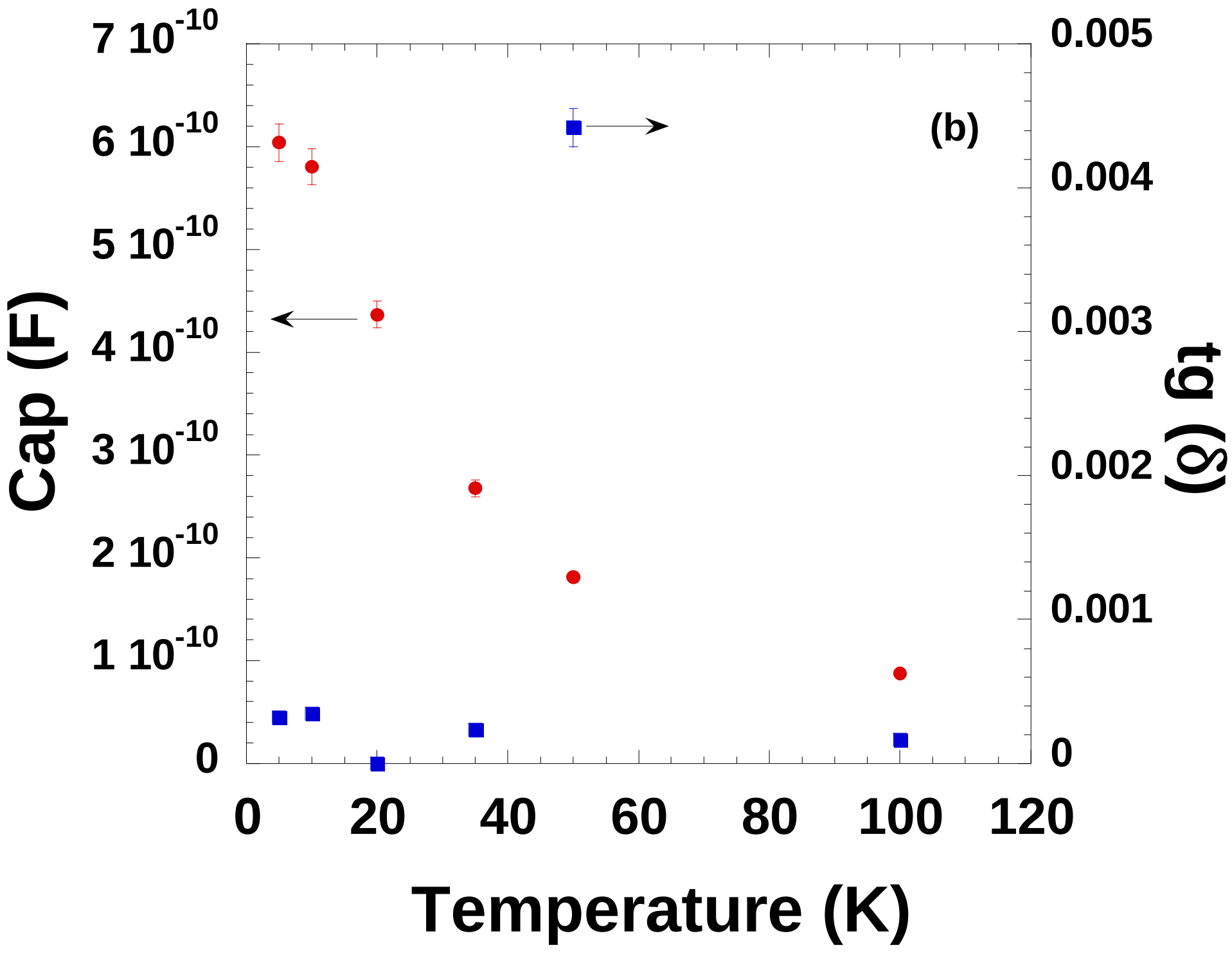

Figure 2

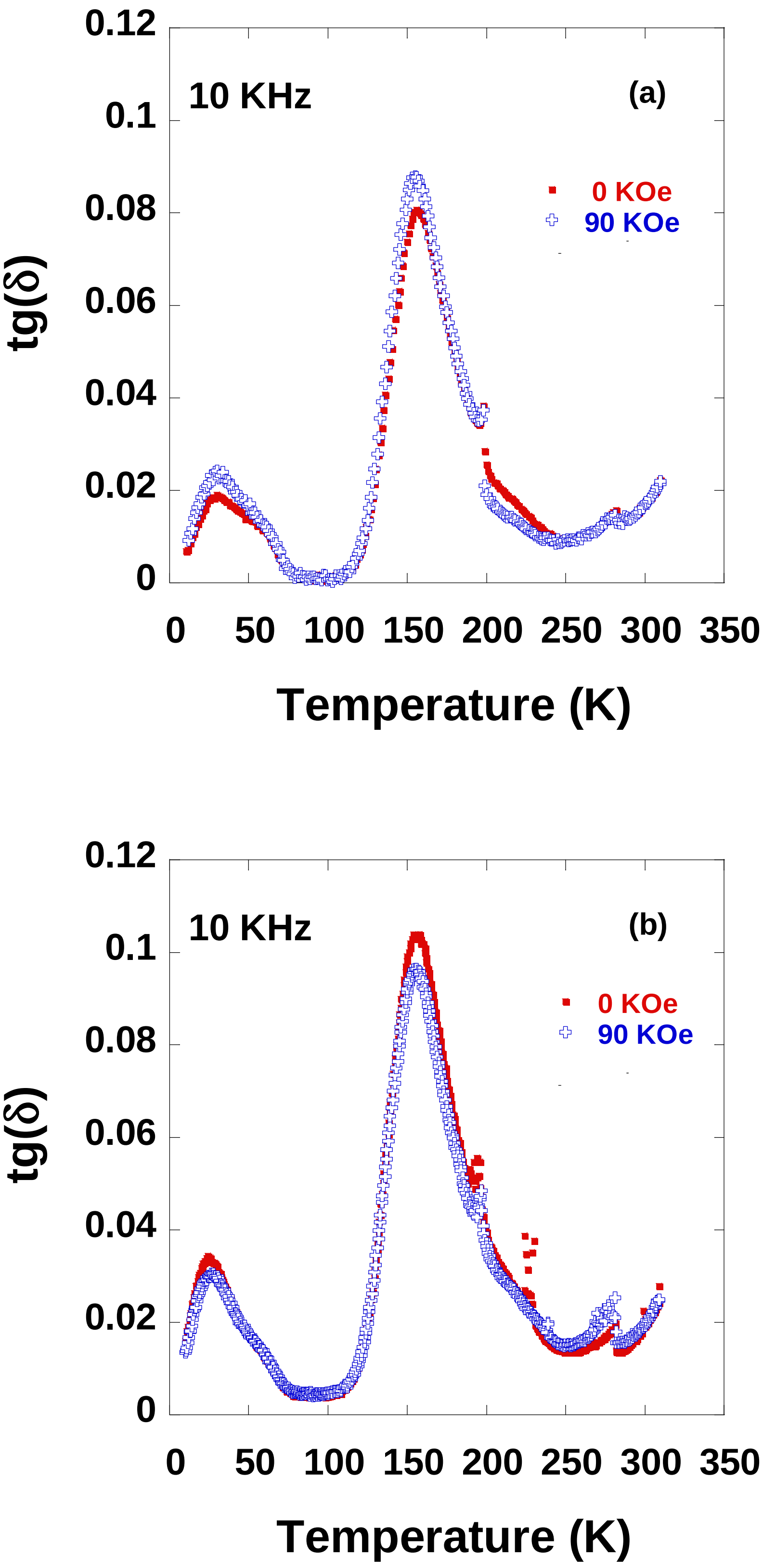

# Figure 3

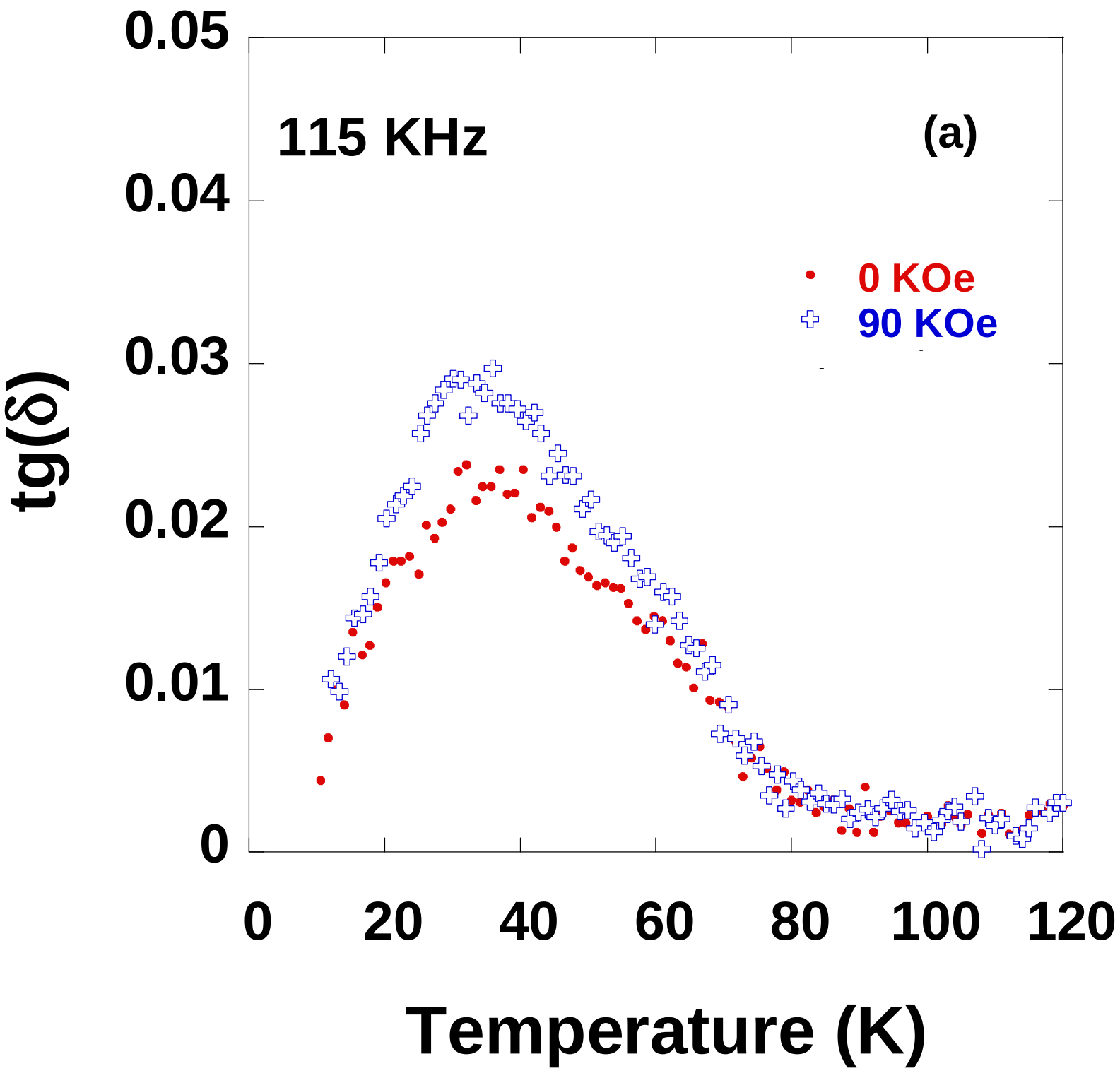

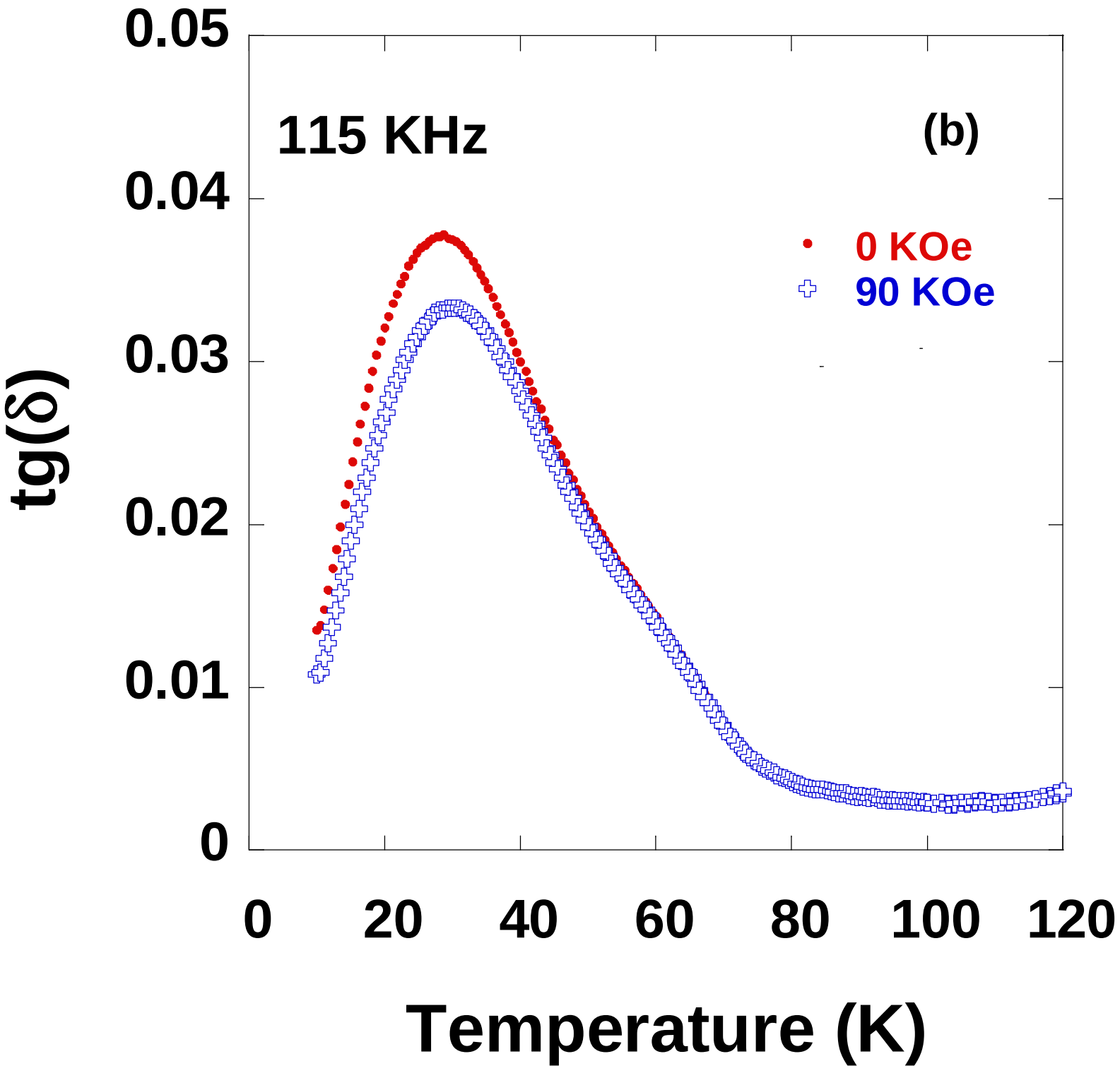

Figure 4

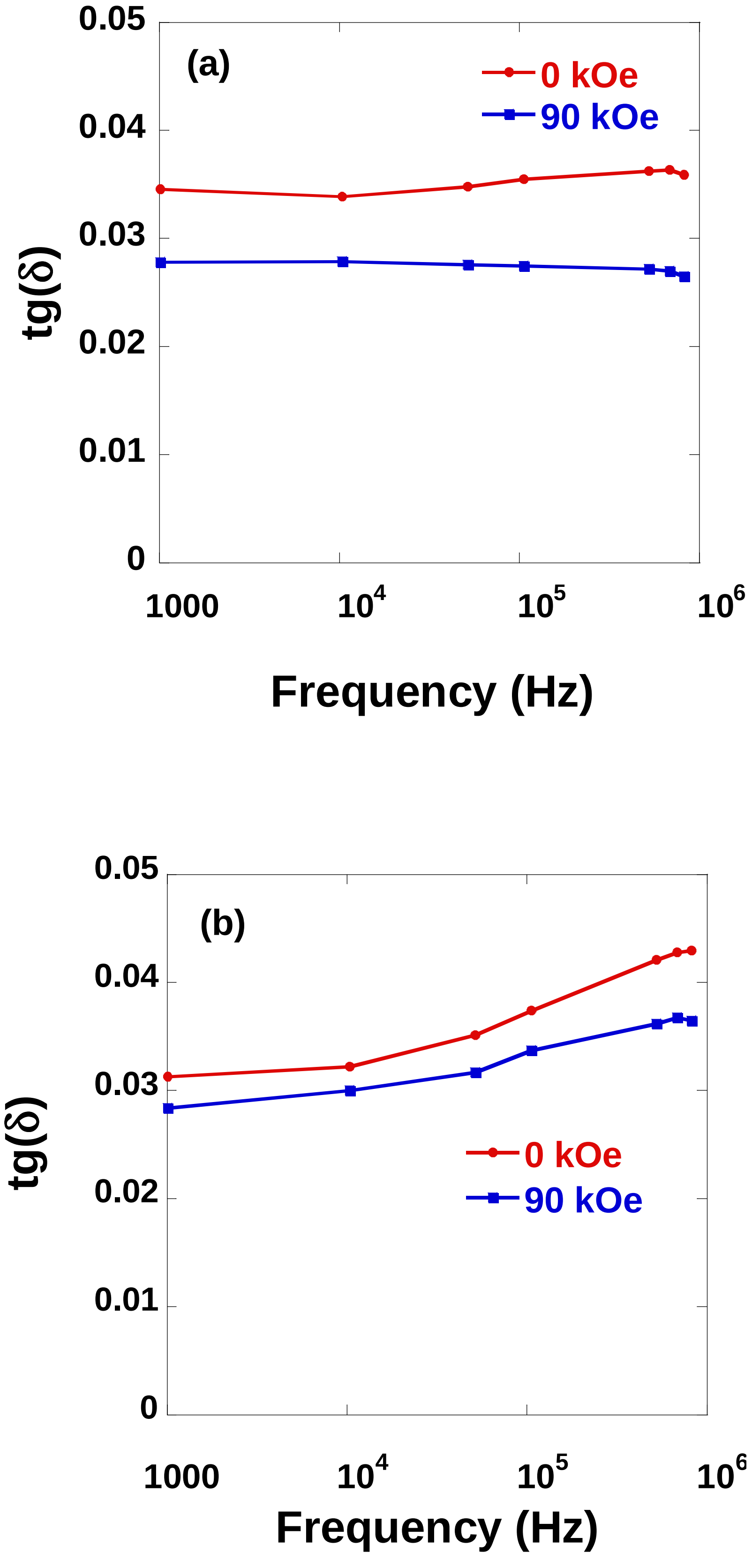

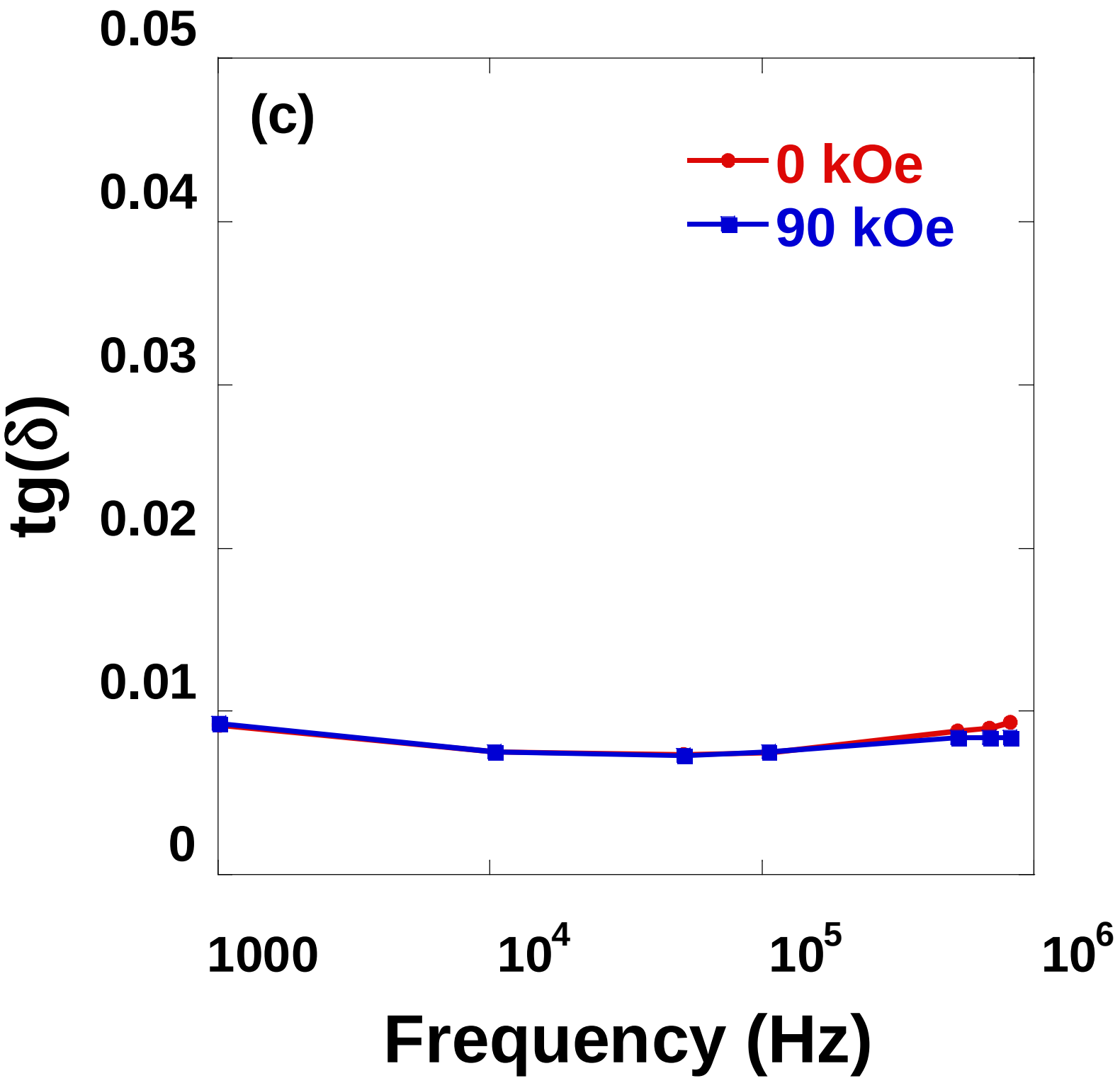

Figure 5

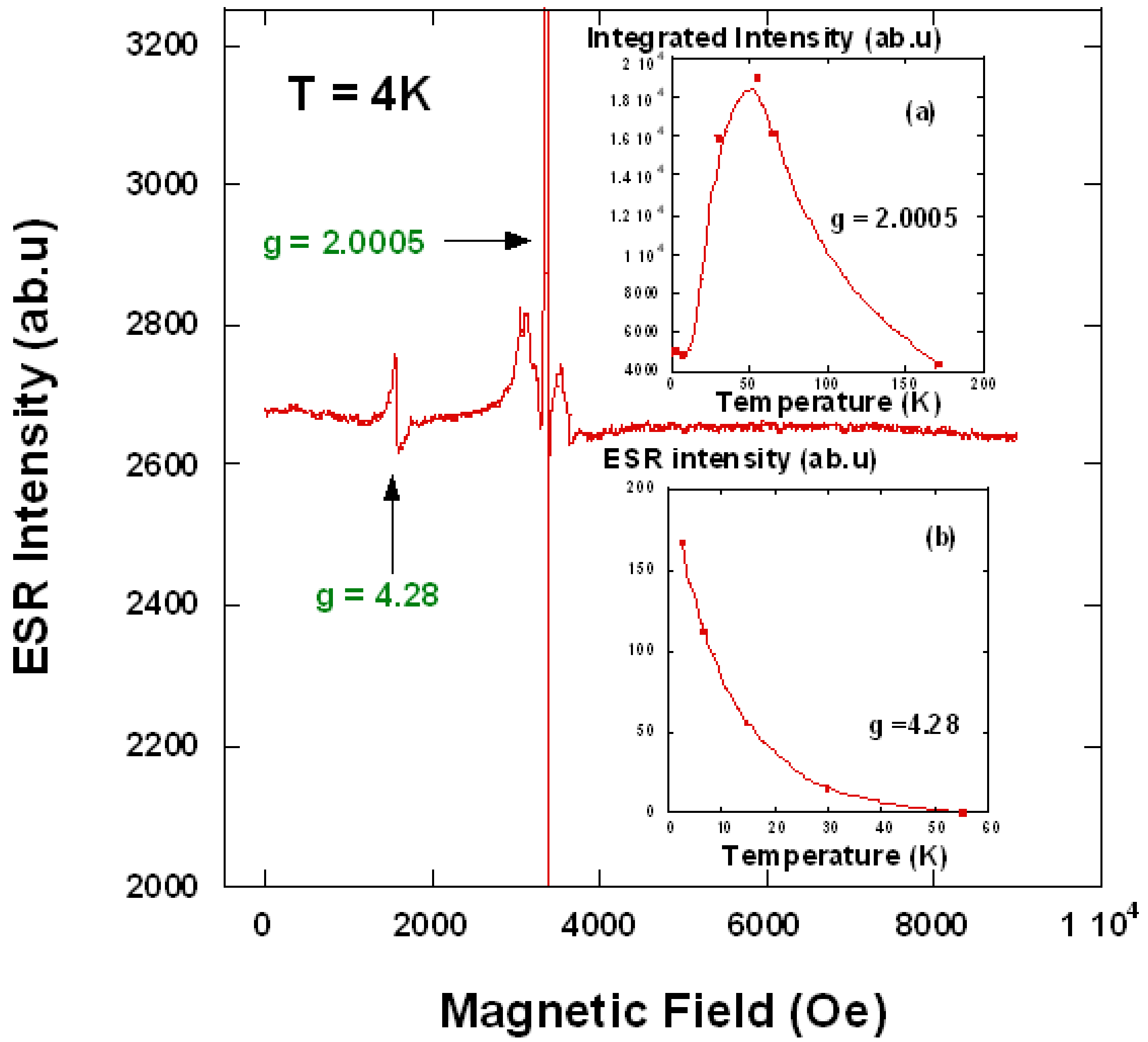